\documentclass[final,5p,times,twocolumn]{elsarticle}

\usepackage{amssymb}
\usepackage{amsmath}
\usepackage{amsthm}
\usepackage{amsfonts}
\usepackage{color}
\usepackage{epstopdf}

\usepackage{subcaption}
\usepackage{graphicx}
\usepackage{bbm}
\usepackage{dsfont}
\usepackage{braket}
\usepackage{hyperref}
\hypersetup{
    colorlinks=true,
    linkcolor=blue,
    filecolor=magenta,      
    urlcolor=cyan,
}
\usepackage{stfloats}

\usepackage{mathtools,etoolbox}
\DeclarePairedDelimiterX{\abs}[1]{\lvert}{\rvert}{\ifblank{#1}{{}\cdot{}}{#1}}

\DeclareMathOperator{\sign}{\operatorname{sign}}

\usepackage{array}

\newcolumntype{x}[1]{%
>{\centering\hspace{0pt}}p{#1}}%

\newcommand{\tnhl}{\tabularnewline\hline}

\usepackage[english]{babel}

\begin{document}

\begin{frontmatter}

% This article is published as part of the Special Issue on "Neuronal Spike Timing: Precision, Reliability, and the Neural Code from 1995-2025"

\title{Quantifying spike train synchrony and directionality: Measures and Applications}

\author[1,2]{Thomas Kreuz} %\orcidlink{0000-0002-9213-8255}}
\ead{thomas.kreuz@cnr.it}

\address[ISC]{Institute for Complex Systems (ISC), National Research Council (CNR), Sesto Fiorentino, Italy}
\address[INFN]{National Institute of Nuclear Physics (INFN), Florence Section, Sesto Fiorentino, Italy}

%%==================================%%
%% Sample for unstructured abstract %%
%%==================================%%

\begin{abstract}
By introducing the twin concepts of reliability and precision along with the corresponding measures, Mainen and Sejnowski's seminal 1995 paper "Reliability of spike timing in neocortical neurons" \cite{Mainen95} paved the way for a new kind of quantitative spike train analysis. In subsequent years a host of new methods was introduced that measured both the synchrony among neuronal spike trains and the directional component, e.g. how activity propogates between neurons. This development culminated with a new class of measures that are both time scale independent and time resolved. These include the two spike train distances ISI- and SPIKE-Distance as well as the coincidence detector SPIKE-Synchronization and its directional companions SPIKE-Order and Spike Train Order. This contribution to the special issue will not only review all of these measures but also include two recently proposed algorithms for latency correction which build on Spike Train Order and aim to optimize the spike time alignment of sparse spike trains with well-defined global spiking events. For the sake of clarity, all these methods will be illustrated on artificially generated data but in each case exemplary applications to real neuronal data will be described as well.\footnote{This article is published as part of the Special Issue on "Neuronal Spike Timing: Precision, Reliability, and the Neural Code from 1995-2025"}.
\end{abstract}

\end{frontmatter}

\section{Introduction}\label{Intro}

Mainen and Sejnowski's $30$ year old paper \cite{Mainen95} celebrated here in this special issue has not only inspired a multitude of experimental and theoretical investigations on a wide range of topics such as the mechanisms of spike generation and the intricacies of neural coding, it also started the development of a plethora of new methods to quantify spike train synchrony.

Measuring the degree of synchrony within a set of spike trains is a common task in two major scenarios. Spike trains are recorded either simultaneously from a population of neurons, or in successive time windows from only one neuron. In this latter scenario, repeated presentation of the same stimulus addresses the reliability of individual neurons (as in \cite{Mainen95}), while different stimuli are used to investigate neural coding and to find the features of the response that provide the optimal discrimination \cite{Chicharro11, QuianQuiroga13}. These two applications are related since for a good clustering performance one needs not only a pronounced discrimination between different stimuli (low inter-stimulus spike train synchrony) but also a large reliability for the same stimulus (high intra-stimulus spike train synchrony).

In \cite{Mainen95} itself, the demonstration of reliable and precise spike initiation in the neocortex was achieved by means of two measures based on the post-stimulus time histogram (PSTH), which, fittingly, were termed reliability and precision. Just a year after the spike train metric by Victor and Purpura \cite{Victor96} was proposed which evaluates the cost needed to transform one spike train into the other, using only certain elementary steps.  This was soon followed by the van Rossum metric \cite{VanRossum01} which measures the Euclidean distance between two spike trains after convolution of the spikes with an exponential function. Other approaches from that time quantify the cross correlation of spike trains after exponential \cite{Haas02} or Gaussian filtering \cite{S_Schreiber03}, or exploit the exponentially weighted distance to the nearest neighbor in the other spike train \cite{Hunter03}.

A commonality to all of these measures is the existence of one parameter that sets the time scale for the analysis. By contrast, in 2007 the parameter-free ISI-Distance was introduced and compared with these existing approaches \cite[the very first citation of which happens to be \cite{Mainen95}]{Kreuz07c}. This was the first of a new class of measures that is not only time scale independent but also time-resolved. In fact, it is this class of measures (which also includes the SPIKE-Distance, SPIKE-Synchronization, SPIKE-Order, and Spike Train Order) as well as algorithms and applications derived from them that is the topic to be reviewed in this contribution to the special issue.

An overview of all the measures and algorithms presented in this article can be found in Table \ref{Table1:Overview}. First, there are the ISI-Distance \cite{Kreuz07c} and the SPIKE-Distance \cite{Kreuz13}, two spike train distances (i.e., measures inverse to synchrony) that focus on instantaneous comparisons of firing rate and spike timing, respectively (Chapter \ref{ISI-SPIKE-Distance}). A complementary family of measures is given by SPIKE-Synchronization \cite{Kreuz15}, a sophisticated coincidence detector that quantifies the level of synchrony from the number of quasi-simultaneous appearances of spikes, and its directional variants, SPIKE-Order and Spike Train Order \cite{Kreuz17} which allow to sort multiple spike trains from leader to follower and to quantify the consistency of the temporal leader-follower relationships for both the original and the optimized sorting (Chapter \ref{SPIKE-Synchronization-Order}). Building on this SPIKE-Synchronization and Spike Train Order framework, in Chapter \ref{Latency-correction} algorithms are presented which perform latency correction, i.e., optimize the spike time alignment of sparse spike trains with well-defined global spiking events, both for events without \cite{Kreuz22} and with overlap \cite{Mariani25}. Finally, a short outlook is given in Chapter \ref{Conclusions}.\footnote{Note that this contribution specifically focuses on one class of measures (time-scale independent and time-resolved) and its applications. It is not meant as a review of all the various methods for spike train synchrony and directionality that have been proposed over the years. For this the interested reader is referred to the two very comprehensive articles by Cutts \& Eglen \cite{Cutts14} and Baroni \& Fulcher \cite{baroni2025synchrony}.}

\begin{table*}[t]
  	\begin{center}
  		\begin{tabular}{| c | x{3cm} | x{3cm} | x{4cm} |} \hline 			  			
  			\textbf{Section} & \textbf{Measure} & \textbf{What is quantified?} & \textbf{Main purpose} \tnhl
  			2.1 & ISI-Distance & Inverse Synchrony (Distance) & Comparison of instantaneous firing rates  \tnhl
  			2.2 & SPIKE-Distance & Inverse Synchrony (Distance) & Comparison of spike timing (but also sensitive to firing rate) \tnhl
  			3.2 & SPIKE-Synchronization & Synchrony & Matching of spikes \tnhl
  			3.3 & SPIKE-Order & Directionality \hspace{5mm} (of spikes) & Identification of leading and following spikes (color-coding) \tnhl
  			3.3 & Spike Train Order & Directionality \hspace{5mm} (of spike trains) & Identification of leading and following spike trains, then sorting \tnhl
  			3.3 & Synfire Indicator & Directionality \hspace{5mm} (of spike trains) & Quantification of the consistency of that sorting \tnhl \hline

			\textbf{Section} &  \textbf{Algorithm} & \textbf{What is done?} & \textbf{Main purpose} \tnhl
  			4.1 & Latency Correction & Maximize spike alignment & Measure "true" synchrony \tnhl
  			4.2 & Latency Correction (Overlap) & Eliminate overlap & Disentangle global events \tnhl

   		\end{tabular}  		
  	\end{center}  	
  	\caption[Table]{Measures and algorithms presented in this article.
  		\label{Table1:Overview}}
\end{table*}

\vspace{0.5cm}

Throughout this article the number of spike trains is denoted by $N$, indices of spike trains by $n$ and $m$, spike indices by $i$ and $j$ and the number of spikes in spike train $n$ by $M_n$. The spike times of spike train $n$ are thus given as $\{t^{(n)}_i\}$ with $i=1 \dots M_n$. Without loss of generality the interval under consideration is defined to last from time $t = 0$ to $t = T$.

\section{ISI-Distance and SPIKE-Distance} \label{ISI-SPIKE-Distance}

The first step in calculating both the ISI-Distance $D_I$ \cite{Kreuz07c} and the SPIKE-Distance $D_S$ \cite{ Kreuz13} is to transform the sequences of discrete spike times into (quasi-)continuous temporal dissimilarity profiles with one value for each time instant. The temporal profile $I(t)$ of the ISI-Distance is derived from the interspike intervals, while for the SPIKE-Distance the profile $S(t)$ is calculated from differences between the spike times of the spike trains.

Useful for both definitions and starting with just two spike trains, to each spike train $n=1,2$ and each time instant $t$ (Fig. \ref{Fig1-Illustration}A) are assigned three piecewise constant quantities, the time of the previous spike
\begin{equation} \label{Eq:Previous-Spike1}
	t^{(n)}_\mathrm{P}(t)=\max(t^{(n)}_i|t^{(n)}_i \leq t),
\end{equation}
the time of the following spike
\begin{equation} \label{Eq:Next-Spike}
	t^{(n)}_\mathrm{F}(t)=\min(t^{(n)}_i | t^{(n)}_i > t),
\end{equation}
and the interspike interval
\begin{equation} \label{Eq:Interspike-Interval}
	x^{(n)}_{\mathrm{ISI}}(t)=t^{(n)}_F(t) - t^{(n)}_P(t).
\end{equation}
The ambiguity regarding the definition of the very first and the very last interspike interval and the special cases of empty spike trains or spike trains with just one spike are dealt with in \cite{Satuvuori17}.

\subsection{ISI dissimilarity profile}\label{ISI-dissimilarity-profile}

\begin{figure}
	\includegraphics[width = \columnwidth]{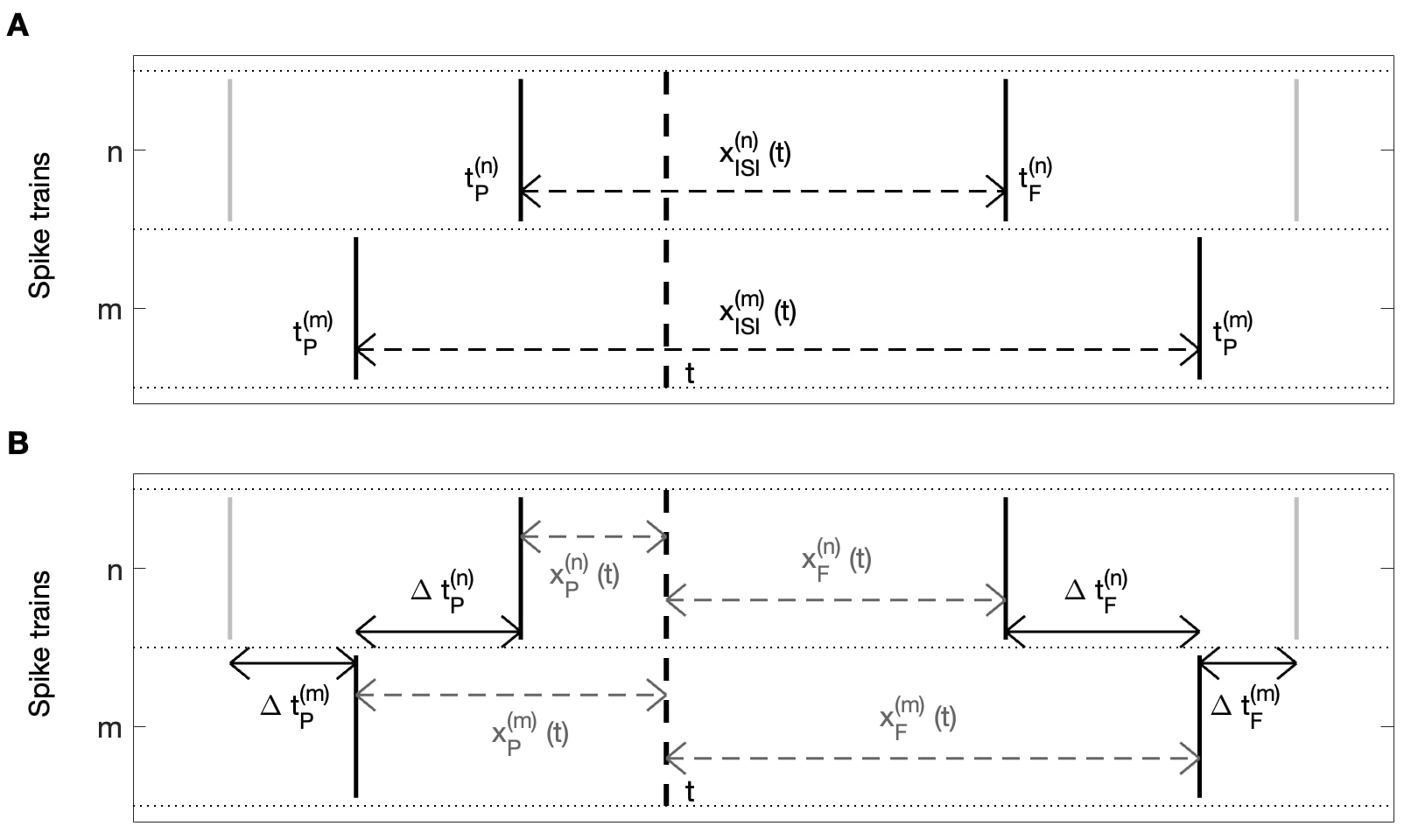}
	\centering
	\caption{Schematic illustration of how the ISI-Distance $D_I$ and the SPIKE-Distance $D_S$ are derived from local quantities around an arbitrary time instant $t$. A. The ISI dissimilarity profile $I (t)$ is calculated from the instantaneous interspike intervals. B. Additional spike-based variables make the SPIKE dissimilarity profile $S (t)$ sensitive to spike timing. Modified from \cite{Kreuz15}.
 \label{Fig1-Illustration}}
\end{figure}

The ISI-Distance \cite{Kreuz07c} and its multivariate extension \cite{Kreuz09} are based on the instantaneous interspike intervals (see Fig. \ref{Fig1-Illustration}A). A time-resolved, symmetric, and time scale adaptive measure of the relative firing rate pattern is obtained by calculating the normalized instantaneous ratio between $x^{(n)}_{\mathrm{ISI}}$ and $x^{(m)}_{\mathrm{ISI}}$ as
\begin{equation} \label{Eq:ISI-profile}
 I(t) = \frac{|x^{(n)}_{\mathrm{ISI}}(t) - x^{(m)}_{\mathrm{ISI}}(t)|}{\max\{x^{(n)}_{\mathrm{ISI}}(t),x^{(m)}_{\mathrm{ISI}}(t)\}}.
\end{equation}
%
% #############################################################################
% ################## Figure 1: Schematics ISI-SPIKE-Sync ######################
% #############################################################################
%

\begin{figure}[!ht]
	\includegraphics[width=\columnwidth]{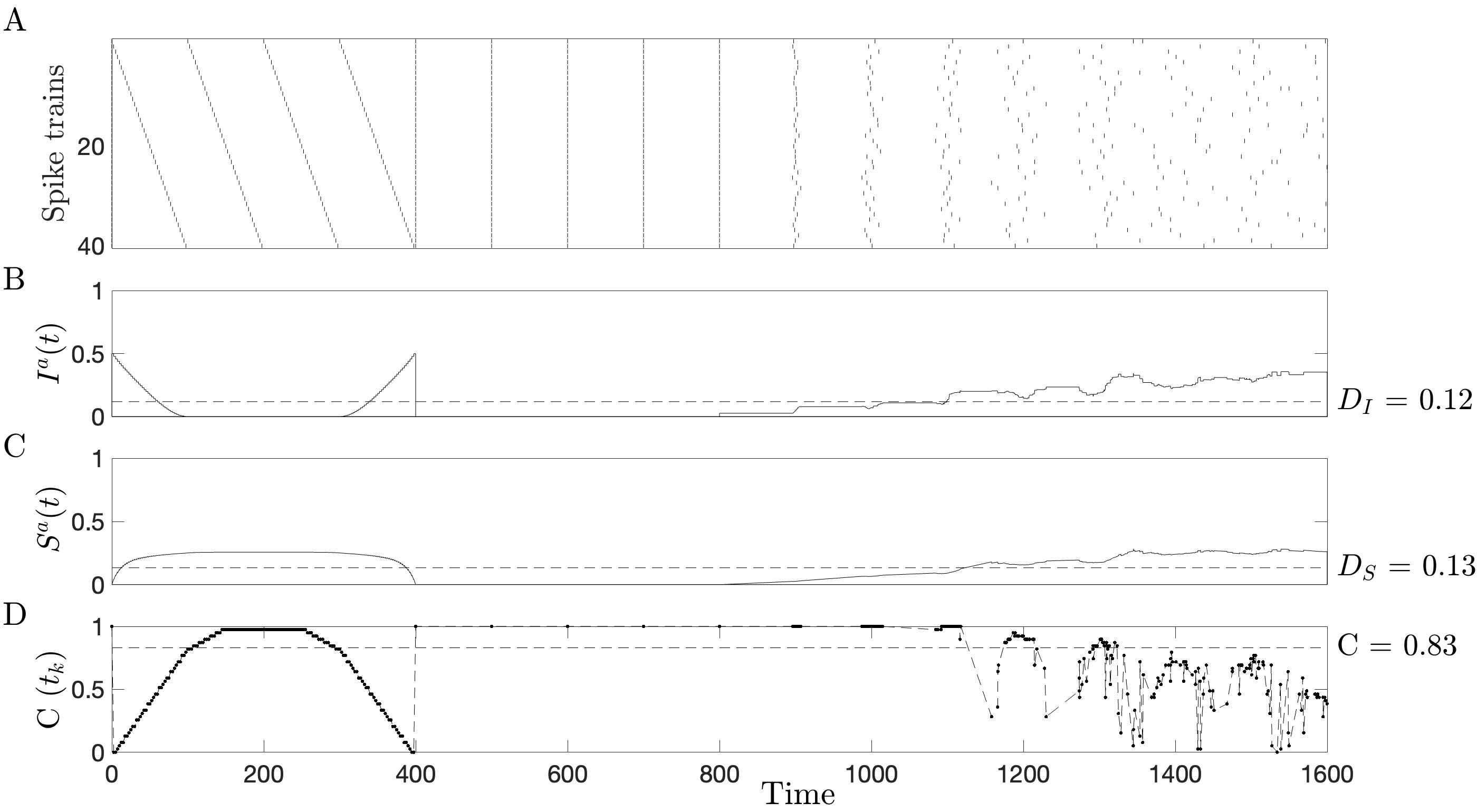}
	\caption{Artificial example dataset of $40$ spike trains (A) and the corresponding profiles for the ISI-Distance (B), the SPIKE-Distance (C), and SPIKE-Synchronization (D), as well as their average values, calculated according to Eqs. \ref{Eq:Multivariate-Distance-2} (ISI- and SPIKE-Distance) and \ref{Eq:SPIKE-Synchronization-C} (SPIKE-Synchronization). The first half illustrates that only the SPIKE-distance and SPIKE-synchronization are sensitive to spike timing (the ISI-distance is zero both in the middle of a synfire chain and for perfect global events even though the spike timing synchrony is very different). The second half shows a slow transition from perfect synchrony to complete randomness which is reflected by gradual increases in the two dissimilarities and a decrease in the SPIKE-synchronization profile (which initially is more tolerant towards deviations in spike timing than the SPIKE dissimilarity).}
	\label{Fig2-ISI-SPIKE-Sync-Profiles}
\end{figure}

This ISI dissimilarity profile becomes $0$ for identical ISI in the two spike trains, and approaches $1$ whenever one spike train has a much higher firing rate than the other. As the interspike intervals are piecewise constant functions, also the dissimilarity profile is piecewise constant. The ISI dissimilarity profile for an artificial example of $50$ spike trains with global events of slowly varying levels of jitter and different noise levels (see Fig. \ref{Fig2-ISI-SPIKE-Sync-Profiles}A) is shown in Fig. \ref{Fig2-ISI-SPIKE-Sync-Profiles}B.
%
% #############################################################################
% ################## Figure 2: ISI-SPIKE-Sync-Profile #########################
% #############################################################################
%

\subsection{SPIKE dissimilarity profile}\label{SPIKE-dissimilarity-profile}

Since the ISI dissimilarity profile is based on the relative length of simultaneous interspike intervals, it is optimally suited to quantify similarities in firing rate profiles. However, it is not designed to track the type of synchrony that is mediated by spike timing (see also \cite{Satuvuori18a}). This is a particular kind of sensitivity which is not only of theoretical but also of high practical importance since coincidences of spikes occur in many different neuronal circuits such as the visual cortex \cite{Usrey99, Tiesinga08} or the retina \cite{Shlens08}. This led to the development of the SPIKE dissimilarity profile which uniquely combines some of the useful properties of the ISI dissimilarity profile with a specific focus on spike timing. The definition presented here is the one introduced in \cite{Kreuz13}, which improves considerably on the original proposal \cite{Kreuz11}.

To derive the instantaneous dissimilarity profile for the SPIKE-Distance, in the beginning to each time instant four corner spikes are assigned (see Fig. \ref{Fig1-Illustration}B): the preceding spike of spike train $n$, $t_{\mathrm{P}}^{(n)}$, the following spike of spike train $n$, $t_{\mathrm{F}}^{(n)}$, the preceding spike of spike train $m$, $t_{\mathrm{P}}^{(m)}$, and, finally, the following spike of spike train $m$, $t_{\mathrm{F}}^{(m)}$. Each of these four corner spikes can then be attached with the spike time difference to the nearest spike in the other spike train, e.g., for the previous spike of spike train $n$, 
\begin{equation} \label{Eq:Delta-Corner-Spike}
     \Delta t_{\mathrm{P}}^{(n)} (t) = \min_i (| t_{\mathrm{P}}^{(n)} (t) - t_i^{(m)} |)
\end{equation}
and analogously for $t_{\mathrm{F}}^{(n)}$, $t_{\mathrm{P}}^{(m)}$, and $t_{\mathrm{F}}^{(m)}$.

Subsequently, for each spike train separately a locally weighted average is employed such that the difference of the closer spike dominates: The weighting factors are the intervals from the time instant under consideration to its previous and to its following spike, e.g.,
\begin{equation} \label{Eq:Previous-SPIKE-Distance}
     x_{\mathrm {P}}^{(n)} (t) = t - t_{\mathrm {P}}^{(n)} (t)
\end{equation}
and
\begin{equation} \label{Eq:Following-SPIKE-Distance}
     x_{\mathrm {F}}^{(n)} (t) = t_{\mathrm {F}}^{(n)} (t) - t.
\end{equation}

Accordingly, for that spike train the local weighting of the two spike time differences reads:
\begin{equation} \label{Eq:Spike-Dissimilarity-1}
     S_n (t) = \frac{\Delta t_{\mathrm {P}}^{(n)} (t) x_{\mathrm {F}}^{(n)} (t) + \Delta t_{\mathrm {F}}^{(n)} (t) x_{\mathrm {P}}^{(n)} (t)}{x_{\mathrm {ISI}}^{(n)} (t)}.
\end{equation}

Averaging over the contributions of both spike trains and normalizing by the mean interspike interval gives the dissimilarity profile of the rate-independent SPIKE-Distance (which was proposed in \cite{Satuvuori17}): 
\begin{equation} \label{Eq:Spike-Dissimilarity-Intermediate}
     S_{RI} (t) = \frac{S_n (t) + S_m (t)}{x_{\mathrm {ISI}}^{(n)} (t) + x_{\mathrm {ISI}}^{(m)} (t)}.
\end{equation}

This quantity takes into account relative distances within each spike train, but ignores differences in time scale between spike trains. In order to account for differences in firing rate and get these ratios straight, in a last step the contributions from the two spike trains are locally weighted by their instantaneous interspike intervals. This defines the SPIKE dissimilarity profile:
\begin{equation} \label{Eq:Spike-Dissimilarity}
     S (t) = \frac{S_n (t) x_{\mathrm {ISI}}^{(m)} (t) + S_m (t) x_{\mathrm {ISI}}^{(n)} (t)}{\frac{1}{2} \left[ x_{\mathrm {ISI}}^{(n)} (t) + x_{\mathrm {ISI}}^{(m)} (t) \right]^2}.
\end{equation}

Since this dissimilarity profile is obtained from a linear interpolation of piecewise constant quantities, it is itself piecewise linear (with potential discontinuities at the spikes). The SPIKE dissimilarity profile of the same artificial spike train set used in Section \ref{ISI-dissimilarity-profile} is shown in Fig. \ref{Fig2-ISI-SPIKE-Sync-Profiles}C.

\subsection{Similarities and differences}\label{ISI-SPIKE}

% Common properties
Both the ISI- and the SPIKE-Distance are defined as the temporal average of the respective time profile, e.g., for the ISI-Distance,
\begin{equation} \label{Eq:Temporal-Average}
    D_I = \frac{1}{T} \int_0^T dt I (t).
\end{equation}

Also, for both distances, there exist two ways to derive from the bivariate versions the multivariate extension to $N > 2$ spike trains. First it can be obtained by simply averaging over only the upper right triangular part (since both distances are symmetric) of the pairwise distance matrix, here again for the ISI-Distance,
\begin{equation} \label{Eq:Multivariate-Distance-1}
    D_I = \frac{2}{N(N-1)}\sum_{n=1}^{N-1} \sum_{m=n+1}^N D_I^{n,m}.
\end{equation}

In this case the temporal average of Eq. \ref{Eq:Temporal-Average} is followed by the (spatial) average over all pairs of spike trains of Eq. \ref{Eq:Multivariate-Distance-1}. However, these two averages commute, so it is also possible to achieve the same kind of time-resolved profile as in the bivariate case by first calculating the instantaneous average $S^{\mathrm {a}} (t)$ (now for the SPIKE-Distance) over all pairwise instantaneous values $S^{n,m} (t)$:
\begin{equation} \label{Eq:Multivariate-Profile}
    S^{\mathrm {a}} (t) = \frac{2}{N(N-1)}\sum_{n=1}^{N-1} \sum_{m=n+1}^N S^{n,m} (t).
\end{equation}

This time the spatial average is followed by the temporal average
\begin{equation} \label{Eq:Multivariate-Distance-2}
    D_S = \frac{1}{T} \int_0^T dt S^{\mathrm {a}} (t),
\end{equation}
and so Eqs. \ref{Eq:Multivariate-Distance-1} and \ref{Eq:Multivariate-Distance-2} yield the exact same value.

% ISI and SPIKE - Ranges and Zero distance
Both dissimilarity profiles, the piecewise constant $I (t)$ and the piecewise linear $S (t)$, as well as both distances $D_I$ and $D_S$ (calculated according to Eq. \ref{Eq:Temporal-Average}) always stay within the interval $[0, 1]$. For the SPIKE-Distance the limit value $D_S = 0$ is obtained only for perfectly identical spike trains, while for the ISI-distance $D_I = 0$ can also be attained for periodic spike trains with exactly the same period. Further mathematical properties of both distances (including expectation values for Poisson spike trains) are derived in \cite{Mulansky15}.

% ISI and SPIKE - Different representations
Spike trains can be analyzed on many different spatial and temporal scales, accordingly these two time-resolved and pairwise measures of spike train dissimilarity allow for several levels of information extraction \cite{Kreuz15}. In the most detailed representation one instantaneous value is obtained for each pair of spike trains, while the most condensed representation of successive temporal and spatial averaging leads to one single distance value that describes the overall level of synchrony for a spike train set over a given time interval. In between these two extremes are spatial averages (multivariate dissimilarity profiles, see Fig. \ref{Fig2-ISI-SPIKE-Sync-Profiles}, B and C) and temporal averages (pairwise dissimilarity matrices, examples for the SPIKE-Distance are shown in Fig. \ref{Fig3-SPIKE-Matrices}). A movie version of these matrices can be found in the Supplemental Material of \cite{Kreuz13} and on Youtube (\url{https://www.youtube.com/watch?v=Po_YxuHW_ZY}).
%
% #############################################################################
% ################## Figure 3: SPIKE-matrices #################################
% #############################################################################
%
\begin{figure*}[!ht]
	\includegraphics[width=\textwidth]{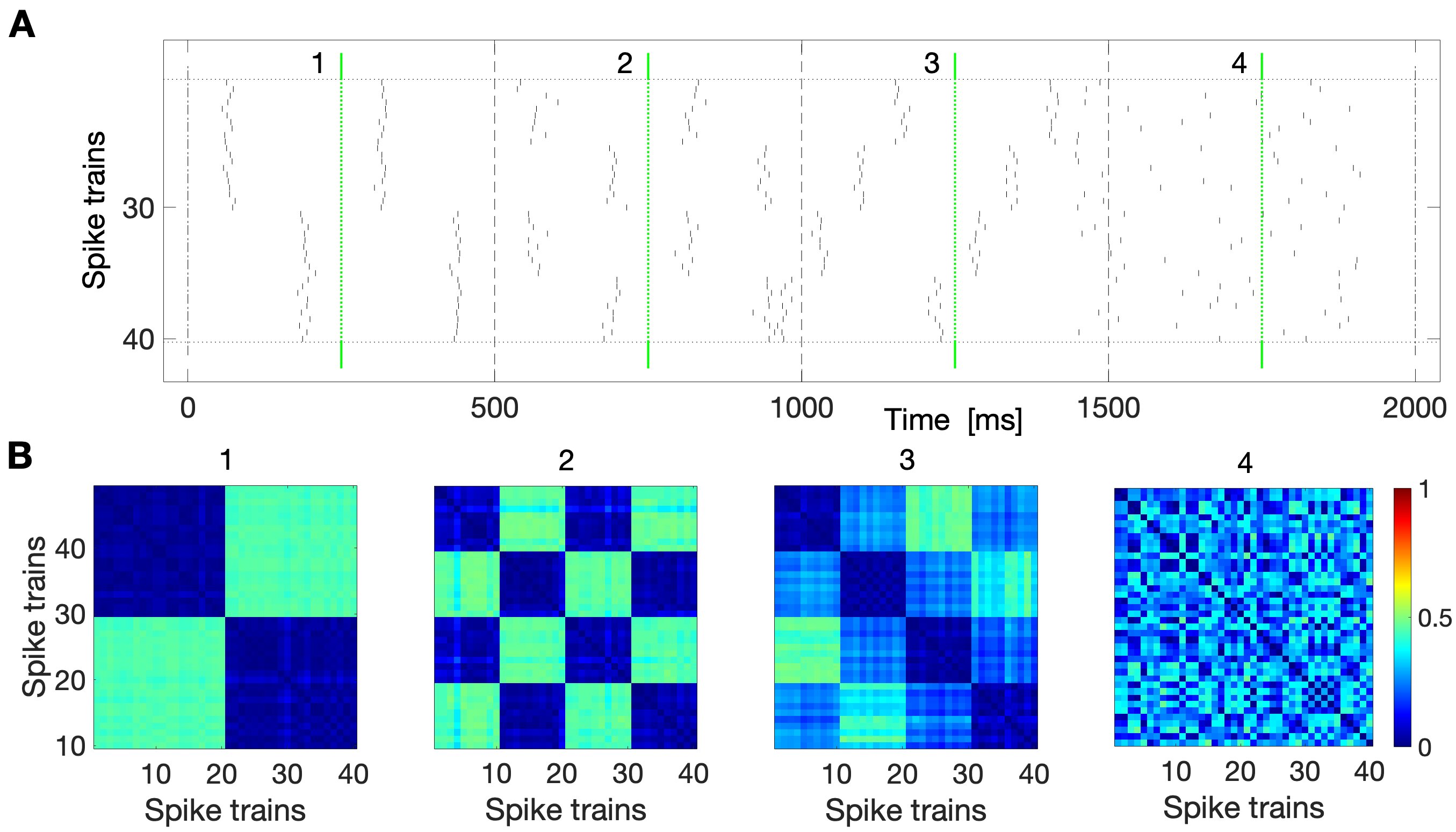}	
	\vspace{0.1cm}
	\caption{Instantaneous clustering for artificially generated spike trains (A) whose clustering behavior changes every $500$ ms: from two different variants of two clusters via four clusters to random spiking. B. The matrices of pairwise instantaneous SPIKE dissimilarity values for the time instants marked by the green lines in A faithfully reflect the clustering at the respective moment. Modified from \cite{Kreuz13}.}
	\label{Fig3-SPIKE-Matrices}
\end{figure*}

\vspace{0.5cm}
\noindent \large{\textbf{ISI-Distance - Example applications}}

Since its proposal in 2007, the ISI-Distance has been applied to electrophyiological data close to 100 times (For a constantly updated lists of such studies for all the measures dealt with here please refer to \url{https://www.thomaskreuz.org/publications/isi-spike-articles}). Here are a few examples: Recently, the ISI-Distance was used as an important feature for simulation-based inference (SBI), a machine learning approach that automatically estimates parameters that replicate the activity of "human induced pluripotent stem cells" (hiPSCs)-derived neuronal networks on multi-electrode arrays (MEAs) \cite{doorn2025automated}. Other studies classified Alzheimer's disease phenotype based on hippocampal electrophysiology \cite{moradi2023early}, assessed the performance of biologically-inspired image processing in computational retina models \cite{melanitis2019biologically}, and performed spike train synchrony analysis of neuronal cultures \cite{lama2018spike}.

\vspace{0.5cm} \noindent \large{\textbf{SPIKE-Distance - Example applications}}

Even though the SPIKE-Distance was proposed six years after the ISI-Distance, it has already surpassed its counterpart in usage (which could possibly indicate again that spike timing matters, but of course there might be other factors that have a bearing on the decision to prefer one measure over the other). The most recent of the more than $100$ applications shows that motor cortex stimulation increases the variability of single-unit spike configuration in parkinsonian but not in normal rats \cite{lee2025motor}. Other studies evaluate the coordinated activity in human induced pluripotent stem cells (iPSCs) derived neuron-astrocyte co-cultures \cite{goshi2025direct}, or perform a non-parametric physiologixcal classification of retinal ganglion cells in the mouse retina \cite{jouty2018non}. Interestingly, the SPIKE-Distance, like all the other measures, is often used in many different fields and contexts outside of neuroscience, for example to assess the reproducibility of eyeblink timing during formula car driving \cite{nishizono2023highly}.

On the other hand, the rate-independent SPIKE-Distance has recently been employed to quantify neural discrimination of different frequency tones in a rat model of Fragile X Syndrome, a leading inherited cause of autism spectrum disorders (ASD) \cite{gauthier2025altered}. It has also helped to dissect the contributions of parvalbumin neurons towards rate and timing-based coding in complex auditory scenes and to explore their ability to reduce cortical noise \cite{nocon2023parvalbumin}.

\section{SPIKE-Synchronization and SPIKE-Order}\label{SPIKE-Synchronization-Order}

\subsection{Spike matching via adaptive coincidence detection} \label{Coincidence-Detection} 

% #############################################################################
% ################## Figure 4: Coincidence Detection ##########################
% #############################################################################
%
\begin{figure}[!ht]
	\begin{center}	
	\includegraphics[width=\columnwidth]{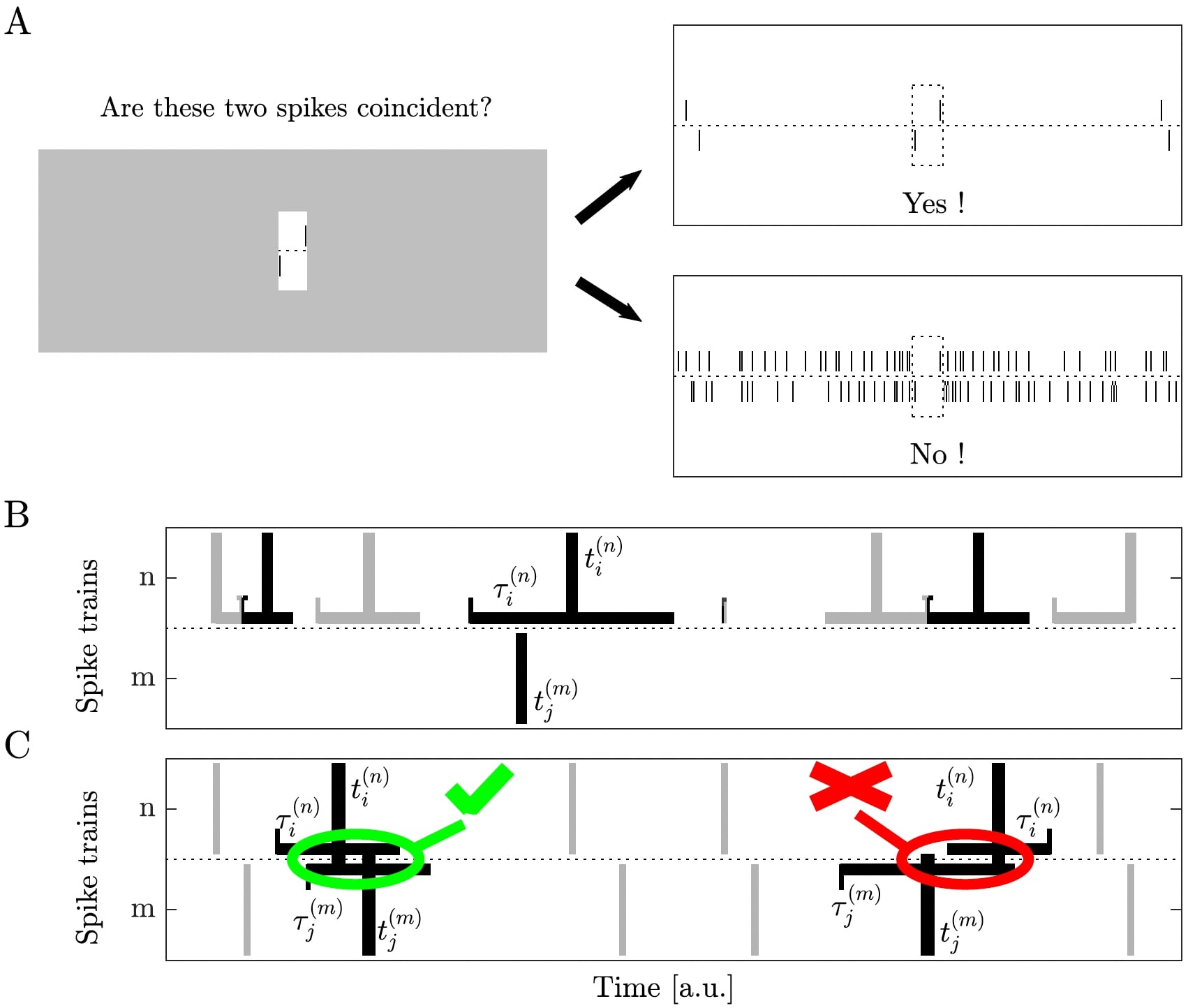}
	\caption{A. Motivation for adaptive coincidence detection. Depending on local context the same two spikes (left) can appear as coincident (right, top) or as non-coincident (right, bottom).
B/C. Illustration of adaptive coincidence detection.
The first step (B) assigns to each spike $t_i^{(n)}$ of spike train $n$ a potential coincidence window that does not overlap with any other coincidence window:
$\tau_i^{(n)} = \min \{t_{i+1}^{(n)} - t_i^{(n)}, t_i^{(n)} - t_{i-1}^{(n)}\}/2$.
Thus any spike from spike train $m$ can at most be coincident with one spike from spike train $n$. Short vertical lines mark the times right in the middle between two spikes. For better visibility spikes and their coincidence windows are shown in alternating bright and dark color.
In the same way (C) a coincidence window
$\tau_j^{(m)} = \min \{t_{j+1}^{(m)} - t_j^{(m)}, t_j^{(m)} - t_{j-1}^{(m)}\}/2$
is defined for spike $t_j^{(m)}$ from spike train $m$. For two spikes to be coincident they have to be in each other's coincidence window which means that their absolute time difference has to be smaller than $\tau_{ij}=\min \{\tau_i^{(n)}, \tau_j^{(m)}\}$ (which is equivalent to Eq. \ref{Eq:Coincidence-MaxDist}). In this example the two spikes on the left are coincident, whereas the two spikes on the right are not. Modified from \cite{Kreuz17}.}
	\label{Fig2-Coincidence-Detection}
	\end{center}
\end{figure}

As Fig. \ref{Fig2-Coincidence-Detection}A illustrates, in general it is basically impossible to judge whether two spikes are coincident or not without taking the local context into account. To overcome this problem, \cite{QuianQuiroga02b} proposed an adaptive coincidence detection which is scale- and thus parameter-free since the minimum time lag $\tau^{(n,m)}_{ij}$ up to which two spikes $t_i^{(n)}$ and $t_j^{(m)}$ from different spike trains are considered to be synchronous is adapted to the local firing rates (higher firing rates lead to smaller coincidence windows):
\begin{equation} \label{Eq:Coincidence-MaxDist}
 \begin{aligned}
    \tau^{(n,m)}_{ij} = \min \{&t_{i+1}^{(n)} - t_i^{(n)}, t_i^{(n)} - t_{i-1}^{(n)},\\
    &t_{j+1}^{(m)} - t_j^{(m)}, t_j^{(m)} - t_{j-1}^{(m)}\}/2.
 \end{aligned}
\end{equation}

Starting with a pair of spikes from two different spike trains (the minimal case for which there can be a coincidence or not) the adaptive coincidence criterion can then be applied in a multivariate context \cite{Kreuz15} by defining for each spike $i$ of spike train $n$ a coincidence indicator (which considers all spikes $j$ of spike train $m$):
\begin{equation} \label{Eq:Coincidence-Indicator}
	C_i^{(n,m)}=\begin{cases}
		1 & {\rm if}  \min_j (|t_i^{(n)} - t_j^{(m)}|) < 
		\tau_{ij}^{(n,m)} \cr
		0 & {\rm otherwise.}
	\end{cases}
\end{equation}

Due to the minimum function and the "$<$" (instead of "$\leq$") any spike can at most be coincident with one spike (the nearest one) in the other spike train (Fig. \ref{Fig2-Coincidence-Detection}B) and thereby an unambiguous spike matching is guaranteed. The coincidence indicator $C_i^{(n,m)}$ is either $1$ or $0$ depending on whether the spike $i$ of spike train $n$ is part of a coincidence with any spike of spike train $m$ or not (Fig. \ref{Fig2-Coincidence-Detection}C).

\subsection{SPIKE-Synchronization}\label{SPIKE-Synchronization}

In order to derive an overall measure of spike matching, this adaptive criterion of "closeness in time" is applied to all spikes of a given spike train set. By averaging over all $N-1$ bivariate coincidence indicators involving spike $i$ of spike train $n$, a multivariate normalized coincidence counter is obtained:
\begin{equation} \label{Eq:SPIKE-sync-Multi-Counter}
	C_i^{(n)}=\frac{1}{N-1} \sum_{m \neq n} C_i^{(n,m)}.
\end{equation}

Subsequently, pooling the coincidence counters of the whole spike train set results in a single multivariate SPIKE-Synchronization profile
\begin{equation} \label{Eq:SPIKE-sync-Multi-Profile}
    	\{C(t_k)\} = \bigcup_n \{C_{i(k)}^{(n(k))} \},
\end{equation}
where the spike indices $i(k)$ and the spike train indices $n(k)$ are mapped onto a global spike index $k$.

With $M = \sum_{n=1}^N M_n$ denoting the total number of spikes, the average of this profile yields the SPIKE-Synchronization
\begin{equation} \label{Eq:SPIKE-Synchronization-C}
	C = \begin{cases}
		\frac{1}{M} \sum_{k=1}^M C(t_k) & {\rm if} \  M > 0 \cr
		\ \ \ \ \ \ \ \ \ 1 & {\rm otherwise,}
	\end{cases}
\end{equation}
the overall fraction of coincidences in the whole spike train set \cite{Kreuz15}. SPIKE-Synchronization attains the value $0$ if and only if there are no coincidences at all and reaches the value $1$ if and only if each spike in every spike train has one matching spike in all the other spike trains - or if there are no spikes at all (since common silence can also be considered as perfect synchrony). The profile for the artificial example from Section \ref{ISI-dissimilarity-profile} is shown in Fig. \ref{Fig2-ISI-SPIKE-Sync-Profiles}D, and a discussion of the mathematical properties of SPIKE-Synchronization can again be found in \cite{Mulansky15}.

\vspace{0.5cm} \noindent \large{\textbf{SPIKE-Synchronization - Example applications}}

SPIKE-Synchronization was proposed only in 2015, nevertheless it has already been employed more than $50$ times. In two very recent publications SPIKE-Synchronization has been used to analyze short-term plasticity in excitatory-inhibitory networks \cite{shan2025short} and to identify the optimal point for transitioning from one gait to another in legged robot locomotion \cite{rostro2025enhancing}. In other studies SPIKE-Synchronization has been applied as machine learning feature in order to discriminate the three states of a network of stochastic spiking neurons \cite{bai2024topological} or to show that birds multiplex spectral and temporal visual information via retinal on- and off-channels \cite{seifert2023birds}.

\subsection{SPIKE-Order}\label{SPIKE-Order}
% #############################################################################
% ################## Figure 5: Spike-Train-Order-Motivation ###################
% #############################################################################
%
\begin{figure}[!ht]
%	\begin{center}
	\includegraphics[width=\linewidth]{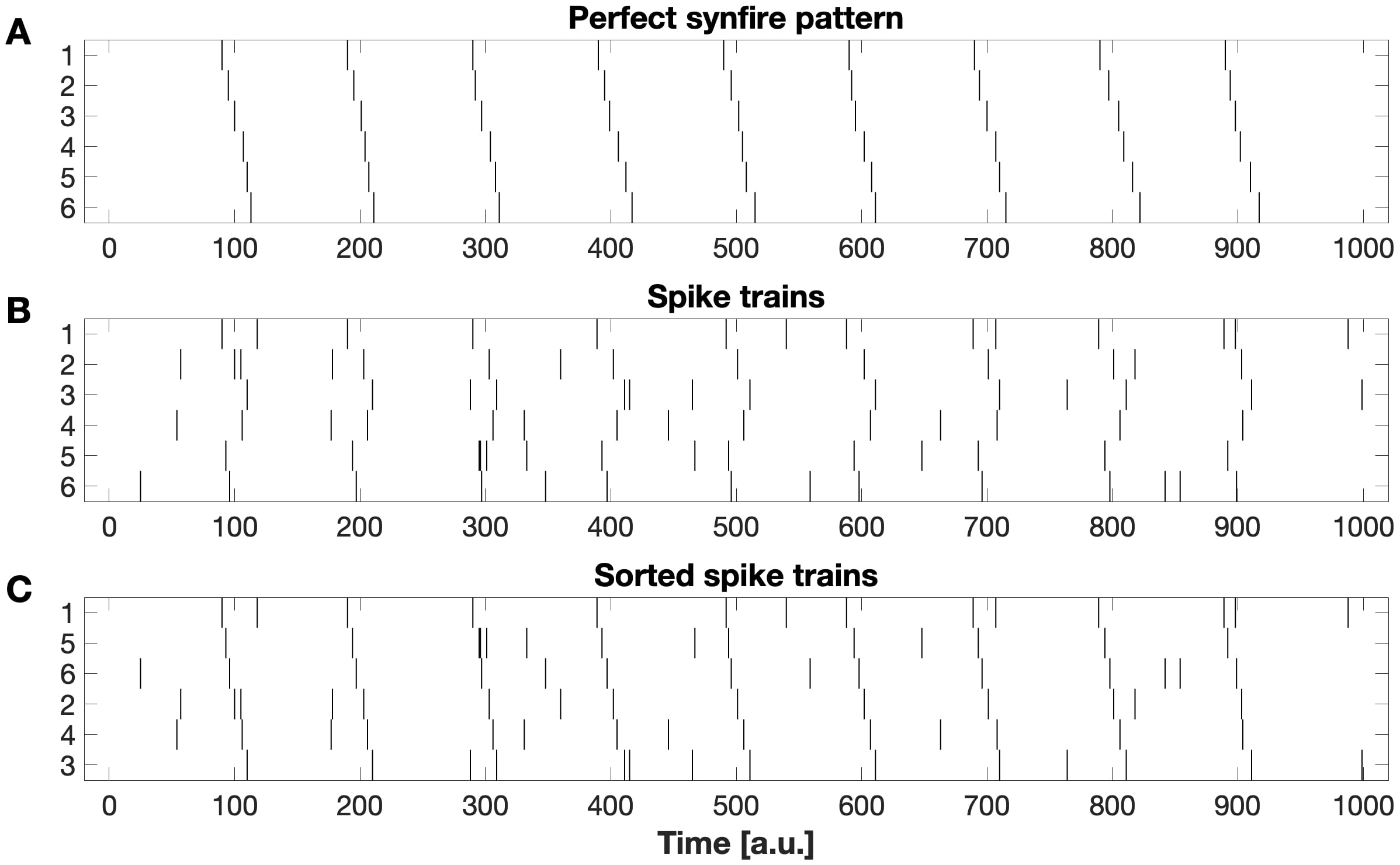}
	\caption{Using the Spike Train Order framework to sort spike trains from leader to follower. A. Perfect synfire pattern. B. Unsorted set of spike trains. C. The same spike trains as in B but now sorted. Modified from \cite{Kreuz17}.}
	\label{Fig5-Spike-Train-Order-Motivation}
%	\end{center}
\end{figure}

% #############################################################################
% ################## Figure 6: Spike-Train-Order ##############################
% #############################################################################
%
\begin{figure*}[!ht]
	\includegraphics[width=\textwidth]{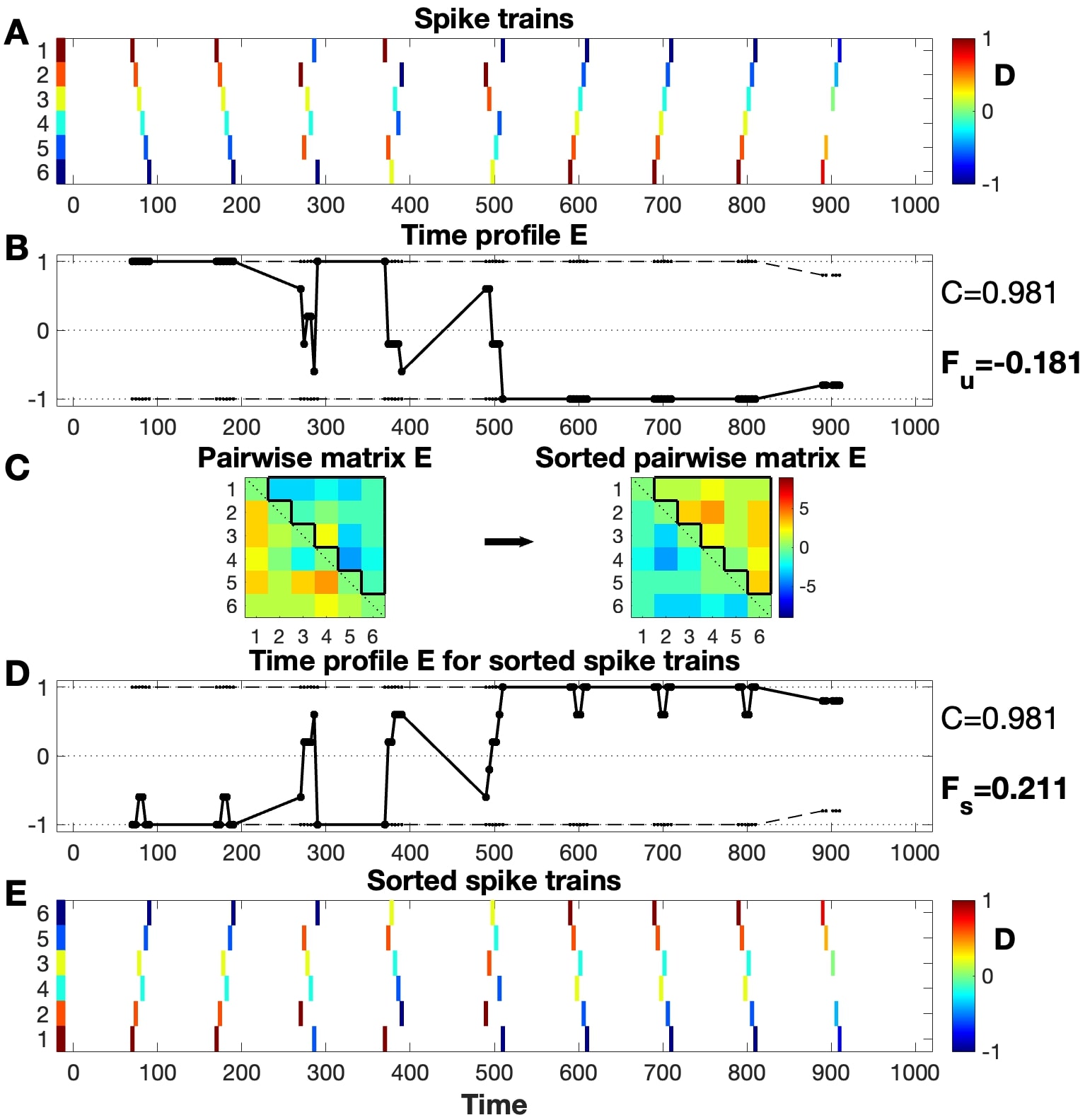}
	\caption{Spike Train Order for an artificial dataset consisting of $6$ spike trains arranged into nine reliable events. The first two events are in order, the last four in inverted order, and for the three events in between the order is random. In the last event one spike is missing. A. Unsorted spike trains with the spikes color-coded according to the value of the SPIKE-Order $D (t_k)$. B. Spike Train Order profile $E (t_k)$. Events with different firing order can clearly be distinguished. The Synfire Indicator $F_u$ for the \textbf{u}nsorted spike trains is slightly negative reflecting the predominance of the inversely ordered events. C. Pairwise cumulative Spike Train Order matrix $E$ before (left) and after (right) sorting. The optimal order maximizes the upper right triangular matrix $E_<$ (Eq. \ref{Eq:Spike-Train-Order-Halfmatrix}), marked in black. The arrow in between the two matrices indicates the sorting process. D. Spike Train Order profile $E (t_k)$ and its average value, the Synfire Indicator $F_s$ for the \textbf{s}orted spike trains (shown  in subplot E). Modified from \cite{Kreuz17}.}
	\label{Fig6-Spike-Train-Order}
\end{figure*}

Often a set of spike trains repeatedly exhibits well-defined patterns of spatio-temporal propagation where activity first appears at a specific location and then spreads to other areas until potentially becoming a global event. If a set of spike trains exhibits perfectly consistent repetitions of the same global propagation pattern, this is called a \textit{synfire pattern} (for an example see Fig. \ref{Fig5-Spike-Train-Order-Motivation}A).

On the other hand, for any given spike train set containing propagation patterns the question arises whether there is spatiotemporal consistency in these patterns, i.e., to what extent do they resemble a synfire pattern, and are there spike trains that consistently lead global events and others that invariably follow these leaders?

The symmetric measure SPIKE-Synchronization (Section \ref{SPIKE-Synchronization}) is invariant to which of the two spikes in a coincidence pair is the leader and which is the follower. To take into account the temporal order of the spikes, the directional measures SPIKE-Order and Spike Train Order \cite{Kreuz17} are needed. Building on SPIKE-Synchronization, the Spike Train Order framework allows to sort the spike trains from leader to follower (compare Figs. \ref{Fig5-Spike-Train-Order-Motivation}B and \ref{Fig5-Spike-Train-Order-Motivation}C) and to evaluate the consistency of the preferred order via a measure called the Synfire Indicator. The application of the whole procedure to a rather simple example dataset can be traced in Fig. \ref{Fig6-Spike-Train-Order}.

First, the spike in spike train $m$ that matches spike $i$ in spike train $n$ is identified as
\begin{equation} \label{Eq:Matching-Spike}
	j' = \arg \min_j ( |t_i^{(n)} - t_j^{(m)}| ).
\end{equation}

Subsequently, the bivariate anti-symmetric SPIKE-Order 
\begin{eqnarray} \label{Eq:SPIKE-Order-Spike-D}
	D_i^{(n,m)} & = & C_i^{(n,m)} \cdot \sign (t_{j'}^{(m)} - t_i^{(n)}) \nonumber \\
	D_{j'}^{(m,n)} & = & C_{j'}^{(m,n)} \cdot \sign (t_i^{(n)} - t_{j'}^{(m)}) = - D_i^{(n,m)},
\end{eqnarray}
assigns to each spike either a $+1$ or a $-1$ depending on whether it is leading or following the coincident spike in the other spike train (and accordingly for that coincident spike). If the two spikes occur at exactly the same time, they both get a zero.

Since SPIKE-Order distinguishes leading and following spikes, it is used to color-code the individual spikes on a leader-to-follower scale (see, e.g., Fig. \ref{Fig6-Spike-Train-Order}A). However, its profile is invariant under exchange of spike trains and thus it looks the same for all events, independent of the order of the spikes within the event. Moreover, averaging over the $D$-profile values of all spikes (which is equivalent to averaging over all coincidences) necessarily leads to a mean value of $0$, since in each coincidence for every leading spike $(+1)$ there has to be a following spike $(-1)$.

Spike Train Order $E$ is similar to SPIKE-Order $D$ but there are two important differences: First, this value depends on the order of the spike trains (and not on the order of the spikes), and second, it is symmetric, so both spikes are assigned the same value:
\begin{equation}  \label{Eq:Spike-Train-Order-E-Spike-1}
 E_i^{(n,m)} = C_i^{(n,m)} \cdot
 				\begin{cases}
 					\sign (t_{j'}^{(m)} - t_i^{(n)})\quad\text{if}\quad n<m\\
                		\sign (t_i^{(n)} - t_{j'}^{(m)})\quad\text{if}\quad n>m
              	\end{cases}
\end{equation}
and
\begin{equation} \label{Eq:Spike-Train-Order-E-Spike-2}
	E_{j'}^{(m,n)} = E_i^{(n,m)}.
\end{equation}
In particular, Spike Train Order assigns to both spikes a $+1$ ($-1$) in case the two spikes are in the correct (wrong) order, i.e., the spike from the spike train with the lower spike train index is the leader (the follower). Once more the value $0$ is obtained when the time of the two coincident spikes is absolutely identical (but also when there is no coincident spike in the other spike train).

The multivariate profile $E(t_k)$, again derived according to Eq. \ref{Eq:SPIKE-sync-Multi-Profile}, is also normalized between $1$ and $-1$ and the extreme values belong to a completely coincident event with all spikes emitted in the correct (incorrect) order from first (last) to last (first) spike train, respectively (compare the first two versus the last four events in Fig. \ref{Fig6-Spike-Train-Order}B). The value $0$ is obtained either if a spike is not part of any coincidence or if the order is such that correctly and incorrectly ordered spike train pairs cancel each other.

By construction (Eqs. \ref{Eq:SPIKE-Order-Spike-D} and \ref{Eq:Spike-Train-Order-E-Spike-1}), $C_k$ is an upper bound for the absolute value of both $D_k$ and $E_k$. In contrast to the SPIKE-Order profile $D_k$, for the Spike Train Order profile $E_k$ it does make sense to calculate the average value. In fact, this is the first way to define the Synfire Indicator (which is described in more detail just below):
\begin{equation} \label{Eq:Synfire-Indicator-F1}
	F = \frac{1}{M} \sum_{k=1}^M E(t_k).
\end{equation}

The important task of sorting the spike trains from leader to follower can be achieved via the cumulative and anti-symmetric Spike Train Order matrix
\begin{equation} \label{Eq:Spike-Train-Order-Matrix}
    E^{(n,m)} = \sum_i E_i^{(n,m)}
\end{equation}
which quantifies the temporal relationship between the spikes in spike trains $n$ and $m$. If $E^{(n,m)}>0$, spike train $n$ is leading spike train $m$ (on average), while $E^{(n,m)}<0$ implies $m$ is the leading spike train. For a Spike Train Order in line with the synfire property (i.e., exhibiting consistent repetitions of the same global propagation pattern), $E^{(n,m)} > 0$ for all $n<m$ (and accordingly $E^{(n,m)} < 0$ for all $n>m$). Due to the forced anti-symmetry of the matrix there is redundancy of information, so the overall Spike Train Order can simply be derived as the sum over the upper right tridiagonal part of the matrix $E^{(n,m)}$:
\begin{equation} \label{Eq:Spike-Train-Order-Halfmatrix}
 	E_< = \sum_{n<m} E^{(n,m)}.
\end{equation}

Normalizing this cumulative quantity by the total number of possible coincidences yields the second definition of the Synfire Indicator:
\begin{equation} \label{Eq:Synfire-Indicator-F2}
	F = \frac{2 E_<}{(N-1) M}. 
\end{equation}

This definition is equivalent to Eq. \ref{Eq:Synfire-Indicator-F1}. The only difference is that here the temporal summation over the profile is performed before and not after the spatial summation over spike train pairs.

The Synfire Indicator quantifies to what degree coinciding spike pairs with correct order prevail over coinciding spike pairs with incorrect order. It is normalized between $-1$ and $1$ and the value $1$ corresponds to a perfect synfire pattern while the value $-1$ is obtained for a perfectly inverse synfire pattern (i.e., one where the last spike train leads and the first spike train follows). It thus becomes clear that this quantity is a function of the spike train order (which can be denoted as $\varphi(n)$) and that maximizing the Synfire Indicator $F_\varphi$ (starting from the initial (\textbf{u}nsorted) order of the spike trains $\varphi_\textbf{u}$) finds the sorting of the spike trains from leader to follower such that the \textbf{s}orted set $\varphi_\textbf{s}$ comes as close as possible to a perfect synfire pattern \cite{Kreuz17}:
\begin{equation} \label{Eq:Sorted-Order}
	\varphi_s: F_{\varphi_s} = \max_\varphi \{F_\varphi\} = F_s.
\end{equation}
 
In \cite{Kreuz17} the sorting is performed using simulated annealing with switching any two spike train as the elementary operation and the Synfire Indicator as the cost function to be maximized. Unlike the unsorted Synfire Indicator $F_{\varphi_u} = F_u$, the optimized Synfire Indicator $F_s$ can only attain values between $0$ and $1$ (for any negative value simply inverting the spike train order makes it positive, and that is even before the actual optimizing). However, from Eq. \ref{Eq:Spike-Train-Order-E-Spike-1} it follows that for any given dataset $F$ can never be higher than the SPIKE-Synchronization $C$ (Eq. \ref{Eq:SPIKE-Synchronization-C}). The maximum value $F = 1$ is only attained when the spike train set can be sorted into a perfect synfire pattern.

As can be appreciated in Fig. \ref{Fig6-Spike-Train-Order}, according to its two alternative definitions, the maximization of the Synfire Indicator is expressed in both the normalized sum of the upper right half of the pairwise cumulative Spike Train Order matrix (Eq. \ref{Eq:Synfire-Indicator-F2}, Fig. \ref{Fig6-Spike-Train-Order}C) and in the average value of the Spike Train Order profile (Eq. \ref{Eq:Synfire-Indicator-F1}, Fig. \ref{Fig6-Spike-Train-Order}D). Along with this, the spike trains (Fig. \ref{Fig6-Spike-Train-Order}E) are now sorted such that the first spike trains have predominantly high values (red) and the last spike trains predominantly low values (blue) of the SPIKE-Order.

Fig. \ref{Fig6-Spike-Train-Order} also illustrates that the results of the complete analysis contain several levels of information. Time-resolved (local) information is represented in the coloring of the spikes (according to the SPIKE-Order $D$) and in the profile of Spike Train Order $E$. Each element of the Spike Train Order matrix characterizes the leader-follower relationship between two spike trains at a time. The Synfire Indicator $F$ characterizes the closeness of the whole dataset to a synfire pattern, both for the unsorted ($F_u$) and for the sorted ($F_s$) spike trains. The sorted order of the spike trains is a very important result in itself since it identifies the leading and the following spike trains. Finally, as an important last step in the analysis it is highly recommended to evaluate the statistical significance of the optimized Synfire Indicator $F_s$ using a set of carefully constructed spike train surrogates (not shown here but fully explained and visualized in \cite{Kreuz17}).

\vspace{0.5cm} \noindent \large{\textbf{SPIKE-Order - Example applications}}

When originally proposed in \cite{Kreuz17}, the Spike Train Order framework was applied to evaluate the consistency of the leader-follower relationships in datasets from two very different fields, neuroscience (giant depolarized potentials in mice slices) and climatology (El Ni\~no sea surface temperature recordings). Later studies used the same framework to demonstrate the reproducibility of activation sequences in children with refractory epilepsy \cite{tomlinson2019reproducibility} and to perform a spatiotemporal analysis of cortical activity obtained by wide-field calcium images in mice before and after stroke \cite{Cecchini21}.

\section{Latency correction}\label{Latency-correction}

\subsection{Latency correction without overlap}\label{Latency-correction-without-overlap}

When estimating synchrony within a spike train set, systemic delays such as the ones in the synfire pattern of Fig. \ref{Fig5-Spike-Train-Order-Motivation}A are a hindrance, since usually the real question is how would the synchrony look like if there was no latency. For example, in the context of neuronal coding, before estimating the reliability of the neuronal responses upon repeated presentation of a stimulus, it would be best to first get rid of any variations in onset latency. Similarly, if the aim is to quantify the faithfulness in the propagation of activity from one neuron to another, this also should be done only after the removal of the propagation delays. The process of eliminating such systematic delays is called latency correction, and the "true" level of synchrony after such realigning of the spike trains is usually higher than the original synchrony. An algorithm for such a latency correction has recently been proposed in \cite{Kreuz22}.

The crucial step is to go beyond SPIKE-Synchronization and Spike Train Order and to not only use their notions of coincidence and order but also take into account the actual temporal intervals between matching spikes. Therefore, after spike matching via adaptive coincidence detection (Eqs. \ref{Eq:Coincidence-MaxDist} and \ref{Eq:Coincidence-Indicator}) the time difference between any matched pair of spikes is calculated as
\begin{equation}
	\delta^{(n,m)}_i = t^{(n)}_i - t^{(m)}_{j'},
\end{equation}
where $j'$ again identifies the coincident spike in spike-train $m$ that matches spike $i$ in spike train $n$ according to Eq. \ref{Eq:Matching-Spike}.

Averaging over all the matched spikes for pair $n,m$ of spike trains
\begin{equation} \label{Eq:Average-Spike-Time-Difference}
	\delta^{(n,m)} =  \frac{1}{\sum_i C_i^{(n,m)}} \sum_i C_i^{(n,m)} \delta_i^{(n,m)}
\end{equation}
yields the antisymmetric $N \times N$ \textit{spike time difference matrix} (STDM) which estimates the pairwise latencies between all spike trains. Similarly, from the same pairwise averaged spike time differences a symmetric \textit{cost matrix} can be defined as
\begin{equation} \label{Eq:Cost}
	c^{(n,m)} =  \sqrt{\frac{1}{\sum_i C_i^{(n,m)}} \sum_i C_i^{(n,m)} [\delta_i^{(n,m)}]^2},
\end{equation}
which, in contrast to Eq. \ref{Eq:Average-Spike-Time-Difference}, guarantees that the value $0$ is obtained if and only if all matched spike pairs are exactly coincident, i.e., all time differences are $0$. The aim of latency correction then becomes to maximally align the spike trains by minimizing the \textit{cost function} $c$, defined as the average of the upper right triangular part of the cost matrix:
\begin{equation} \label{Eq:Cost-function}
	c = \frac{2}{N(N-1)} \sum^{N}_{n<m} c^{(n,m)}.
\end{equation}

% #############################################################################
% ################## Figure 7: Latency-Correction #############################
% #############################################################################
%
\begin{figure*}[!t]
%	\begin{center}
	\includegraphics[width=\textwidth]{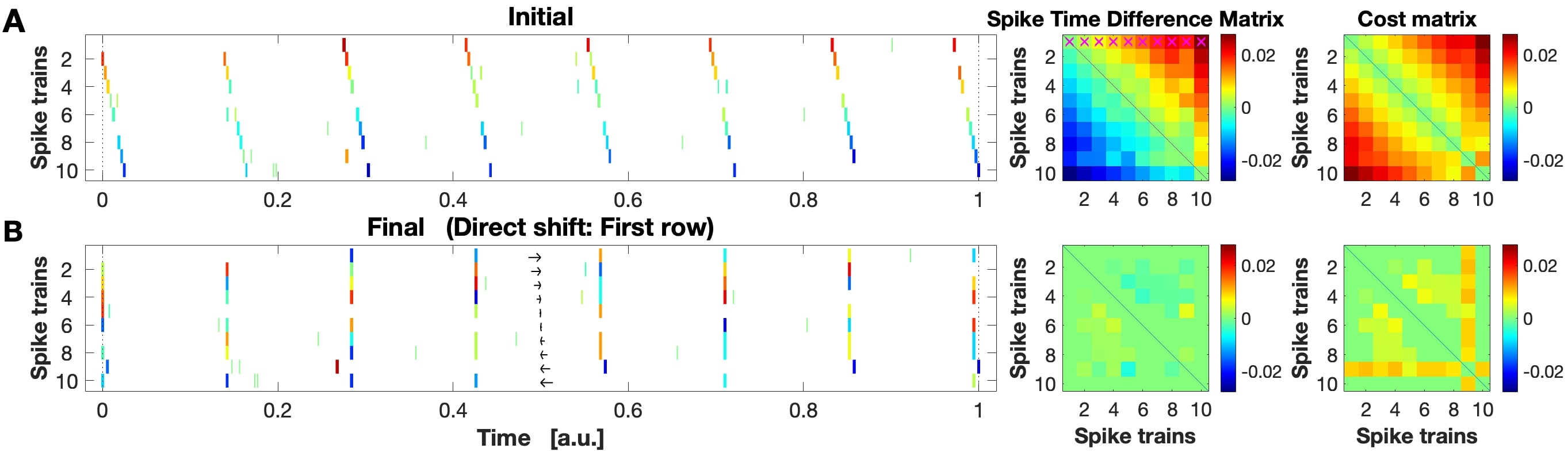}
	\caption{	
	Latency correction performed via first row direct shift on a regularly spaced synfire pattern with both missing and extra spikes as well as some jitter. A. Before and B. After the latency correction. The SPIKE-Order $D$ is color-coded on a scale from $1$ (red, first leader) to $-1$ (blue, last follower). In B the shift performed during the latency correction for each spike train is indicated by arrows. The spike time difference matrix (STDM) and the cost matrix turn from their rather ordered increase away from the diagonal (since initially spikes from more separated spike trains exhibit greater time separation) to very low values everywhere (because the corrected spike trains are almost perfectly aligned). The elements of the STDM used in the direct shift are marked by red crosses. Modified from \cite{Kreuz22}.}
	\label{Fig7-Latency-Correction}
%	\end{center}
\end{figure*}

In \cite{Kreuz22} two latency correction algorithms were proposed. The first algorithm, direct shift, is simple and fast since it takes into account only a minimal part ($N - 1$ values) of the cost matrix. In the \textit{Row Direct Shift} variant these are the values from one row (typically the first, which means that the first spike train is used as reference), while for the \textit{First Diagonal Direct Shift} variant the first upper diagonal (the difference between neighboring spike trains) of the cost matrix is used to calculate the cumulative differences to the first spike train. The correction is performed by shifting the spike trains such that the corresponding matrix elements are set to $0$ and the hope is that this way also the other $(N-1)*(N-2)/2$ matrix elements are minimized.

In Fig. \ref{Fig7-Latency-Correction} on the right hand side the STDM and the cost matrix are shown before and after the latency correction (done using the first row direct shift) for a regularly spaced synfire pattern which was modified by a few missing and extra spikes as well as some jitter. This modification reflects the typical uncertainty inherent in many neuronal recordings, in particular in cases where the spiking data is inferred from calcium imaging. For the uncorrected synfire pattern (Fig. \ref{Fig7-Latency-Correction}A), the further apart two spike trains, the larger the intervals between their matching spikes, and accordingly the values of the STDM and the cost matrix tend to increase with the distance from the diagonal (which itself is necessarily $0$). After the latency correction (Fig. \ref{Fig7-Latency-Correction}B) both matrices approach $0$ everywhere reflecting a much improved alignment of the spike trains, but because of the remaining jitter (which can not be corrected) they do not go all the way down to $0$.

Since in principle the search space of all possible shifts is infinite, the second latency correction algorithm proposed in \cite{Kreuz22} is based on \textit{simulated annealing}, a heuristic approach which employs an iterative Monte Carlo algorithm to minimize the cost function and find the shifts that align the spike trains best. Starting from the initial spike train set and its original cost function $c_{start}$, in each iteration a randomly selected spike train is shifted by a randomly selected time interval (which decreases over time as the cost converges). The cost matrix is updated and the shift is accepted whenever this leads to a decrease of the cost. Unlike greedy algorithms, simulated annealing usually does not get stuck in local minima, since escape is possible because even shifts that lead to an increase in cost are allowed with a certain likelihood (which gets lower and lower as the temperature of the cooling scheme decreases). This iterative procedure continues until the cost function converges towards its final cost $c_{end}$.

For both algorithms the input is a set of $N$ spike trains and there are two major outputs: the end cost $c_{end}$ and the shifts $\vec{s} = [s_1;...;s_N]$ performed in order to get there. In \cite{Kreuz22} it was shown that simulated annealing in general achieves better results (e.g., lower end costs) than direct shifts, however, this improvement comes with a large computational cost of order $N^2$ (whereas for both direct shifts only $N - 1$ additions or subtractions have to be performed).

\subsection{Latency correction with overlap}\label{Latency-correction-with-overlap}

\begin{figure*}[h!]
	\centering
	\includegraphics[width=\textwidth]{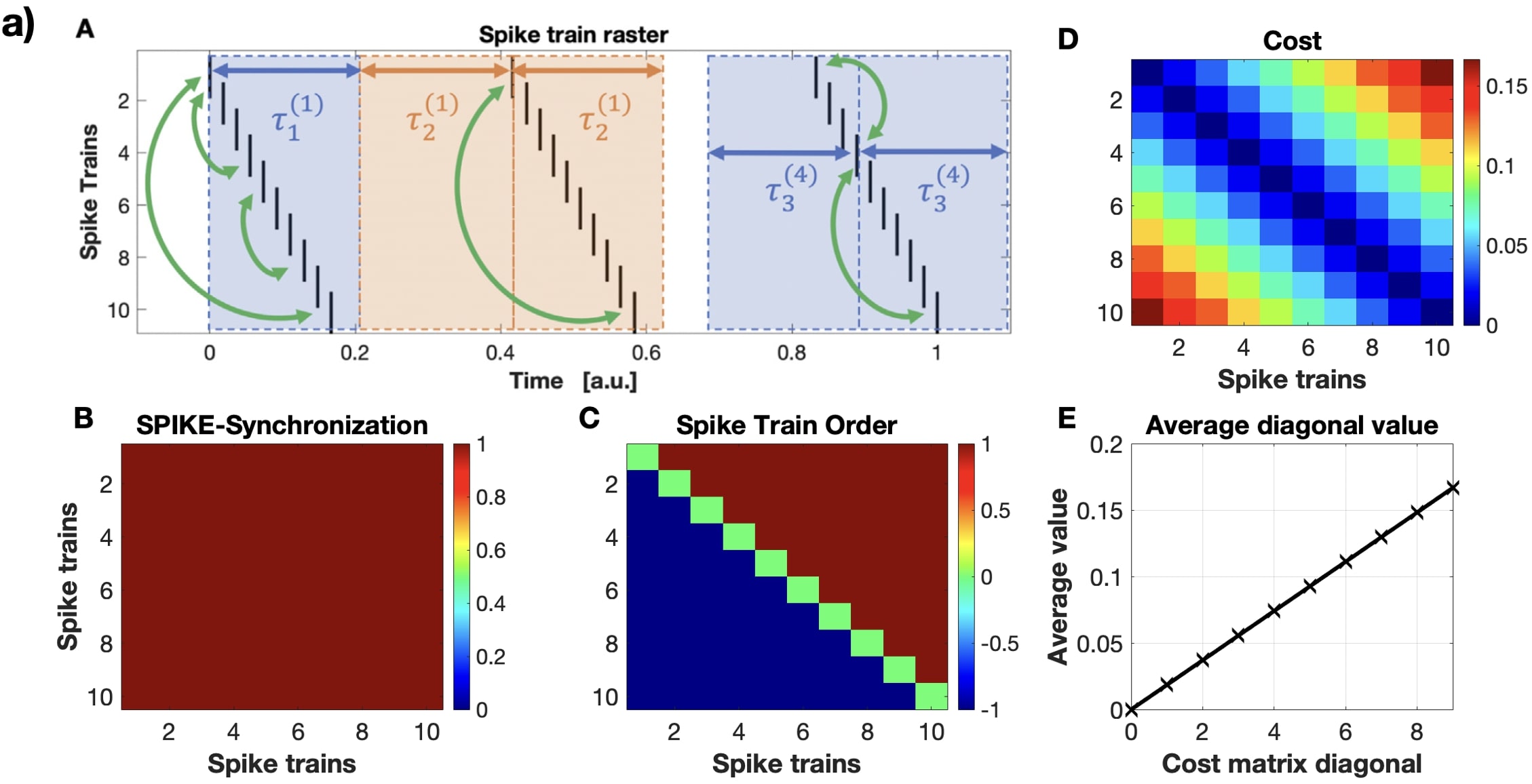}
	\newline
	\vspace{4mm}
	\includegraphics[width=\textwidth]{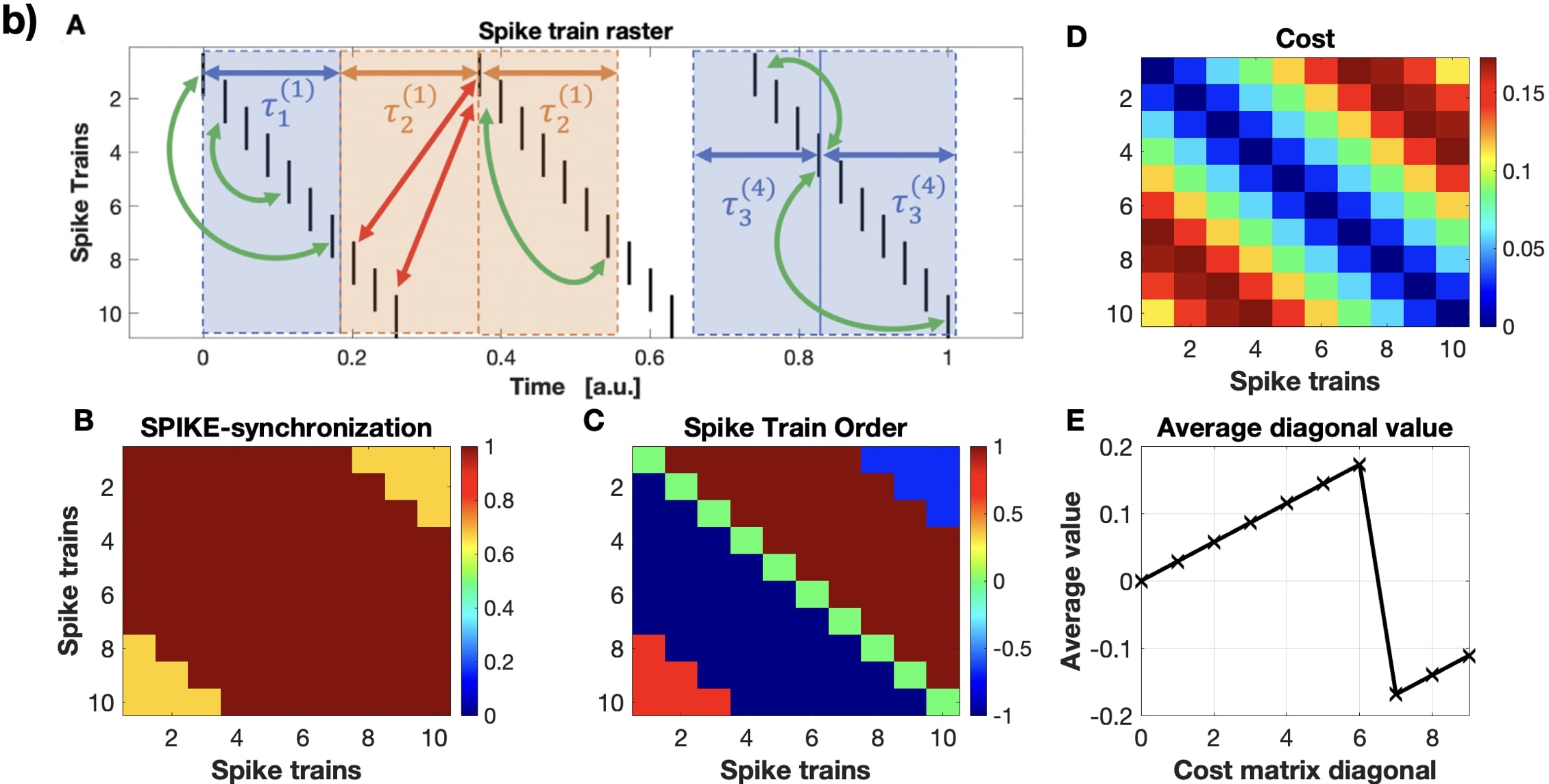}
	%\vspace{.5mm}
	\caption{Illustrating the effect of event overlap using a perfect synfire pattern with three global events without (subpanel a) and with (subpanel b) overlap. Both subpanels follow the same structure: A. Spike train sets. Coincidence windows of the i-th spike in the (n)-th spike train $\tau^{(n)}_i$ are indicated alternately in blue and orange. A few correct and spurious matches are marked by green (both subplots) and red (subplot b only) arrows, respectively. B. SPIKE-Synchronization and C. Spike Train Order matrix. Without overlap (a) both matrices attain maximal values, whereas in (b) overlap results in spurious matches which lead to deviations from $1$ in the off diagonal corners. D. Cost matrix and E. Average diagonal value of the cost matrix. While without overlap (a) the cost increases monotonically with the distance from the diagonal, with overlap (b) the maximum is reached already at an intermediate diagonal. Modified from  \cite{Mariani25}.}
	\label{Fig8-Latency-Correction-with-Overlap}
\end{figure*}

The two algorithms introduced in Section \ref{Latency-correction-without-overlap} work very well as long as the global events are sufficiently separated, for example in the case of a perfect synfire pattern such as the one shown in Fig. \ref{Fig8-Latency-Correction-with-Overlap}a, subplot A. Here each spike of every global event is coincident with all the other spikes of the same event and within each event all spikes are in perfect order. Accordingly, the pairwise matrices of both the SPIKE-Synchronization C (subplot B) and Spike Train Order E (subplot C) attain the maximum value of $1$ everywhere and, in consequence, the same holds true for the overall SPIKE-Synchronization $C$ (Eq. \ref{Eq:SPIKE-Synchronization-C}) and the Synfire Indicator $F$ (Eq. \ref{Eq:Synfire-Indicator-F2}). Finally, the cost matrix (subplot D, Eq. \ref{Eq:Cost}) exhibits a monotonous increase with distance from the main diagonal, as can be quantified by averaging the values of the different diagonals of the cost matrix (subplot E).

However, difficulties start to arise as soon as coincidence windows (Eq. \ref{Eq:Coincidence-MaxDist}) of neighboring events overlap, and these difficulties affect not only SPIKE-Synchronization, Spike Train Order and the Synfire Indicator but also the cost and therefore the latency correction as well. In Fig. \ref{Fig8-Latency-Correction-with-Overlap}b, subplot A, an example is shown in which the intervals between successive spikes from each global event have become so large that the last three spikes from each event are no longer coincident with the first spikes from the same event but rather with the first spikes from the next event. These spurious mismatches lead to diminishing SPIKE-Synchronization (subplot B) and to inconsistencies in the Spike Train Order (subplot C), and consequently SPIKE-Synchronization $C = 0.956$ and Synfire Indicator $F = 0.778$ are both lower than $1$. Also the cost matrix (subplot D) no longer monotonously increases and instead reaches its maximum not in the very last corner (as in Fig. \ref{Fig8-Latency-Correction-with-Overlap}a) but already at an intermediate diagonal (subplot E).
%
% #############################################################################
% ############### Figure 8: Latency-Correction with overlap ###################
% #############################################################################
%

Under normal circumstances the criterion "closeness in time" used by the adaptive coincidence detection of Eqs. \ref{Eq:Coincidence-MaxDist} and \ref{Eq:Coincidence-Indicator} to match spikes is perfectly reasonable. However, in the case of overlap it leads to the result that some of the spikes matched with each other actually belong to different events. But from the rasterplot one can often clearly recognize that the trailing spikes of an event should belong to that event and not to the next on (or sometimes this can be deduced from external knowledge that is not available to the impartial algorithm). In these cases the mismatches and all the resulting reductions in SPIKE-Synchronization, Synfire Indicator and spike time difference can be considered as spurious. In \cite{Mariani25} the algorithms for latency correction from Section \ref{Latency-correction-without-overlap} were adapted such that in the case of overlapping global events they succeed to "overwrite" the naive event matching and still achieve the desired spike train alignment. 

Both the row and the first diagonal direct shift algorithm introduced in Section \ref{Latency-correction-without-overlap} make use of only a very small part of the STDM matrix and are thus very sensitive to noise in the data: if just one of these $N - 1$ values is not reliable, the resulting shifts can be suboptimal and the overall error large. On the other hand, for simulated annealing the cost function is based on the whole STDM which in the case of overlap also includes the spurious outer parts of the matrix (see Fig. \ref{Fig8-Latency-Correction-with-Overlap}b) and this of course causes its own problems. Therefore, the solutions proposed in \cite{Mariani25} address both of these problems by taking the good parts of each of the two existing approaches while avoiding the drawbacks. First, the new fast direct shift algorithm \textit{Extrapolation} takes into account a larger part of the spike time difference matrix than the two direct shifts (and is thus more robust with respect to noise), but at the same time manages to avoid the outer parts of the STDM that are affected by the spurious matches caused by overlap. Subsequently, also simulated annealing is modified such that is based on a reduced matrix only. Both algorithms use a parameter, the \textit{stop diagonal} $d$, to determine the extent to which the STDM is used.

% ###################################################################
% ######### Figure 9: Overlap: Iterative Latency correction #########
% ###################################################################
%
\begin{figure*}[!t]
    \centering
	\includegraphics[width=\textwidth]{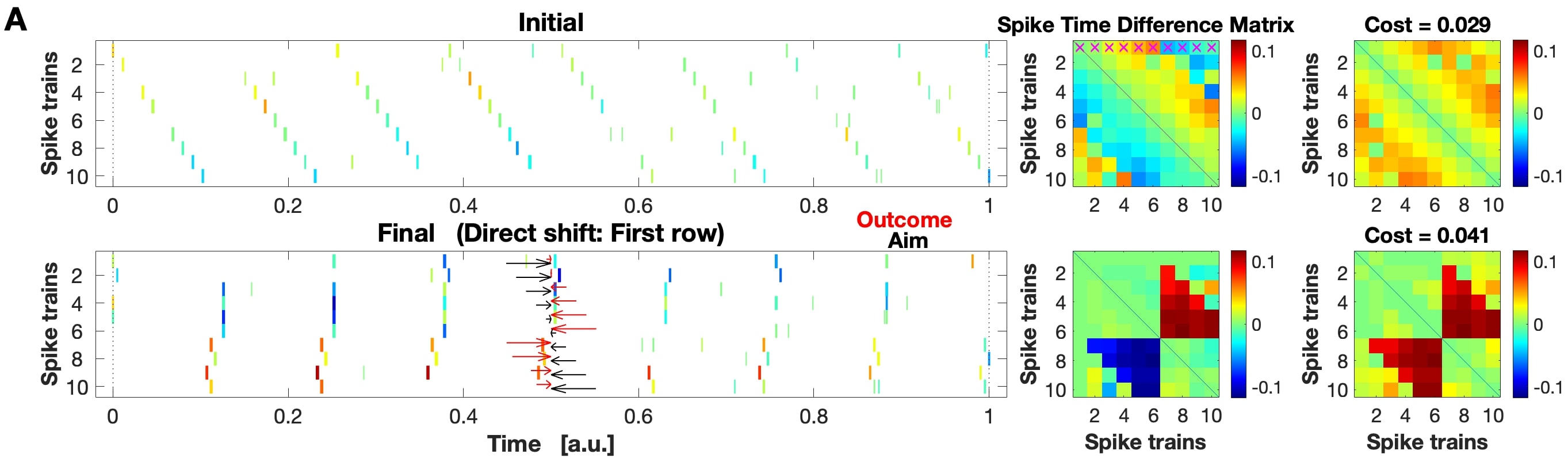}
	\vspace{1mm}	
	\newline
	\includegraphics[width=\textwidth]{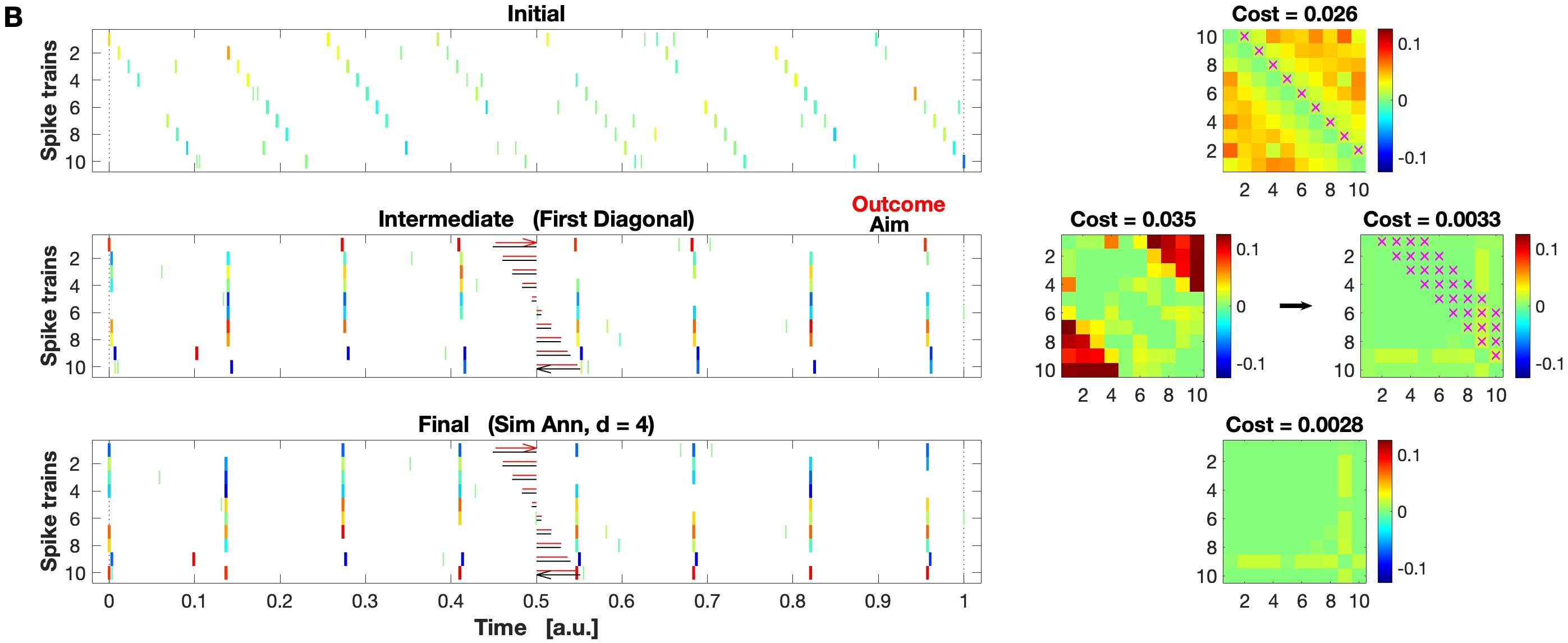}
    \caption{Iterative scheme improves on a simple direct shift when applied to a simulated synfire pattern with overlapping incomplete events and some background spikes. A. The direct shift based on spike time difference values of the first row only (marked by red crosses in the spike time difference matrix) includes incorrect information from the off-diagonal corners of the STDM which are affected by spurious spike matches. It thus tries to align pairs of spikes that are actually mismatched and this erronously breaks up one global event into two parts. B. Iterative scheme composed of a first diagonal direct shift (which avoids the spurious spike matches in the outer parts of the STDM) and a reduced matrix simulated annealing which aims to minimize the cost values up to the fourth diagonal only (again marked by red crosses). The first iteration actually increases the cost value since it is still based on spurious spike matches, but after the rematching (indicated by a black arrow on the right), there is a remarkable decrease in the cost value. In the second iteration the simulated annealing results in a further improvement. In both A and B the shifts that would eliminate the systematic delays of the synfire pattern (Aim) and the shifts obtained by the latency correction algorithms (Outcome) are represented by black and red arrows, respectively. Modified from \cite{Mariani25}.} 
\label{Fig9-Iterative-Latency-Correction}
\end{figure*}

\begin{itemize}

\item Reduced Matrix Direct Shift: Extrapolation

The Extrapolation algorithm builds on a reduced part of the spike time difference matrix and uses the transitivity property to replace the overlap-affected outer parts of the STDM by extrapolating the unaffected inner parts of the matrix:
\begin{equation}
	\delta^{(n, m)} = \sum_{k=m+1}^{n-1} \Bigl[ \delta^{(n, k)} + \delta^{(k, m)} \Bigr]
\end{equation}
for any $n, m$ with $m - n > d$ where $d$ is the stop diagonal parameter. To maintain the anti-symmetry of the STDM, the corresponding elements in the opposite half are updated accordingly:
\begin{equation}
	\delta^{(m, n)} = - \delta^{(n, m)}.
\end{equation}
Conveniently, since each row/column of the STDM contains the shift needed to align with the respective spike train (consistent with a main diagonal of $0$), a simple average over the full STDM immediately results in the invariant shift that uses the median spike train as reference:
\begin{equation}
	s_n =  \frac{1}{N} \sum_{m} \delta^{(m, n)}.
\end{equation}

\item Reduced Matrix Simulated Annealing

For simulated annealing the adaptation to overlapping global events is even more straightforward. Instead of extracting the cost function from the whole upper right triangular part of the STDM, now only the inner overlap-unaffected part of the STDM is utilized, that is, Eq. \ref{Eq:Cost-function} in Section \ref{Latency-correction-without-overlap} is modified such that the sum skips all parts beyond the stop diagonal $d$:
\begin{equation} \label{Eq:Reduced-Cost-function}
	c = \frac{2}{(N-d)(N-d-1)} \sum^{N}_{m-n>d} c^{(n,m)}.
\end{equation}

\end{itemize}

These two algorithms that are both based on reduced matrices already achieve a much better latency correction for cases with overlap, but this can be pushed even further by means of an iterative scheme \cite{Mariani25} in which each iteration consists of two steps: After an initial spike matching via adaptive coincidence detection a first latency correction is performed using a reduced matrix such that it relies only on spike pairs that are not affected by overlap. The aim is to disentangle the different global events and to eliminate as much overlap as possible. Then the second iteration starts again with an adaptive coincidence detection, this time based on spike pairs with a considerably improved alignment and is followed by another latency correction which now typically yields a very increased performance (i.e., lower cost).

This improvement is illustrated in Fig. \ref{Fig9-Iterative-Latency-Correction} on slightly noisy synfire patterns with overlapping events. The "old" first row direct shift of Section \ref{Latency-correction-without-overlap} (Fig. \ref{Fig9-Iterative-Latency-Correction}A) relies on the elements of the first row only and, even worse,  incorporates spurious information from some of the overlap-affected values in the off-diagonal corner of the STDM, which together leads to an excessively high cost value. Instead, in the first iteration of the "new" iterative scheme (Fig. \ref{Fig9-Iterative-Latency-Correction}B) a first diagonal direct shift is used to eliminate the influence of the outer STDM elements that are affected by the spurious spike matches caused by overlap. Since at this moment the calculation of the cost value itself still takes into account outer matrix elements that are based on spurious spike matches, this step actually leads to an increase in the cost (from $0.026$ to $0.035$). However, once the spikes that are now much better aligned within real global events are matched once more, the true cost value turns out to be very low ($0.0033$). The second iteration which employs reduced matrix simulated annealing with stop diagonal $d = 4$ decreases the cost value even further to a final value of $0.0028$.

Finally, in a more extended study on artificially generated data in which we again simulated various degrees of uncertainty in data acquisition (via missing and falsely detected spikes as well as additional jitter) \cite{Mariani25} it could be shown that the Extrapolation direct shift achieves almost the same performance as the reduced matrix simulated annealing, even though it is much faster. For very large datasets this high computational cost of simulated annealing can actually be an exclusion criterion.

\vspace{0.5cm} \noindent \large{\textbf{Latency Correction - Example applications}}

When it was proposed in \cite{Kreuz22}, the regular latency correction algorithm of Section \ref{Latency-correction-without-overlap} (using simulated annealing) was applied to global activation patterns recorded via wide-field calcium imaging in the cortex of mice before and after stroke induced by means of a photothrombotic lesion (these data were first analyzed in \cite{Cecchini21}). In a comparison of three rehabilitation paradigms (motor training for healthy controls, pure motor training for mice with stroke, as well as mice with stroke and additional transient pharmacological inactivation of the controlesional hemisphere), the latter group which is the only one associated with general recovery was also the one that distinguished itself by its lowest end costs. On the other hand, in \cite{Mariani25} the latency correction algorithm for data that contain global events with overlap was tested on single-unit recordings from two medial superior olive neurons of an anaesthetized gerbil during presentations of ten different noisy auditory stimuli \cite{Beiderbeck22}. In all cases the iterative scheme based on the fast extrapolation direct shift (see Section \ref{Latency-correction-with-overlap}) managed to reduce the cost considerable.

\section{Conclusion}\label{Conclusions}

This article reviews a class of measures that are both time scale independent and time-resolved. It presents the basic definitions, shows some illustrative use cases and provides an overview of selected applications within neuroscience, mostly from the fields of neuronal coding and spatio-temporal activity propagation. This is complemented here by a list of a few topics of high importance related to these measures and algorithms that could not be dealt with in sufficient detail, but where the interested reader is directed towards the relevant studies that provide a more in-depth treatment of these topics. For a detailed analysis and discussion of time-scale dependent versus time-scale independent measures of spike train synchrony please refer to \cite{Chicharro11}. The relevance of temporal coding and how spike train distances can address this important problem is discussed in that same paper but also in \cite{Satuvuori18a} which among others provides a table that details which measure is most suitable under which coding hypotheses (this also depends on the firing rate regime). Finally, regarding a statistical analysis and an evaluation of the significance of results the expectation values for Poisson spike trains derived for the three distance / synchrony measures can be found in \cite{Mulansky15} and the spike train surrogate analysis for the SPIKE-order framework in \cite{Kreuz17}. 

Looking into	 the future, there are several promising avenues for methodological advancement: For the SPIKE-Distance and SPIKE-Synchronization (and potentially even SPIKE-Order and Spike Train Order) one could try to find a way to add event magnitudes as weights to individual events, similar to what has been done with the earth mover distance (which is closely related to the Victor-Purpura metric \cite{Victor96, sihn2019spike}). Among others, this could be very interesting in a meta-analysis of epileptic seizure recordings, where the length of the seizure would serve as weight. Comparing results with or without using these weights would provide very valuable information about the significance of the seizure duration, for example in epileptic seizure risk forecasting \cite{leguia2021seizure}. For all the methods and algorithms there is also a need to explore in more detail the effect of unreliability in data acquisition which becomes increasingly relevant, particularly in cases in which larger and larger amounts of spiking activity are inferred from large-scale calcium imaging. In latency correction we actually looked at this \cite{Kreuz22, Mariani25}, but also in this area there are some unresolved issues. For example, it is still necessary to explore in more detail the intertwined relationship between the sorting of spike trains (which is based on latencies as well) and the latency correction itself \cite{Mariani25}.

Last but not least, there are many more potential applications to experimental data: For the two spike train distances (ISI and SPIKE) and SPIKE-synchronization (as well as their adaptive generalizations \cite{Satuvuori17}) these include tracking the neuronal reliability in datasets similar to the one in \cite{Mainen95}, more thorough investigations of stimulus discrimination and clustering in the context of neuronal coding \cite{QuianQuiroga13}, as well as detailed analysis of spatio-temporal propagation patterns in state of the art multi-electrode recordings \cite{schroter2025advances}. On the other hand, the Spike Train Order framework combined with latency correction will allow to judge the faithfulness of activity propagation between different neurons or between different brain areas \cite{Kumar10} or, in the context of neuronal coding, to estimate to what extent the response to repeated presentation of a stimulus is independent of variations in onset latency \cite{levakova2015review}.

While all of these measures and algorithms have been developed within a neuroscientific context, the algorithms are universal and can be applied to discrete datasets in many scientific fields, from climatology (where the measured variable is typically either the temperature or the amount of rainfall) \cite{Kreuz17, sun2018patterns, Conticello20} via network science \cite{mwaffo2018detecting, Bardin19}, social communication \cite{Varni10}, and mobile communication \cite{wang2022identifying} to policy diffusion \cite{grabow2016detecting}.

% Source Codes (Rephrased from 2022 paper)
Optimized implementations of all the measures of spike train synchrony and directionality are available in three free software packages. These are the Matlab graphical user interface SPIKY\footnote[1]{http://www.thomaskreuz.org/source-codes/spiky} \cite{Kreuz15}, the Python library PySpike\footnote[2]{http://mariomulansky.github.io/PySpike} \cite{Mulansky16} and the Matlab command line library cSPIKE\footnote[3]{http://www.thomaskreuz.org/source-codes/cspike}. All of these will soon also include the various algorithms for latency correction as well as a recently proposed algorithm that finds within a larger neuronal population the most discriminative subpopulation \cite{Satuvuori18b}.

\section{Acknowledgements}

A great thanks goes to all the coauthors of the main papers dealt with in this review. Financial support was provided by the National Centre for HPC, Big Data and Quantum Computing - HPC (Centro Nazionale 01 - CN0000013) CUP B93C22000620006 with particular reference to Spoke 8: In Silico Medicine \& Omics Data.

%\bibliography{Kreuz_Bibliography}% common 
\bibliographystyle{elsarticle-num}

\end{document}